\documentclass[aip,jcp,preprint]{revtex4-1}
\usepackage{graphicx}
\usepackage{amssymb}
\usepackage[]{subfigure}
\newlength \figwidth
\begin{document}
\title{Ice polyamorphism in the minimal Mercedes-Benz model of water}
\author{Julyan H. E. Cartwright}
\affiliation{Instituto Andaluz de Ciencias de la Tierra, CSIC--Universidad de Granada, E-18071 Granada, Spain}
\email{julyan.cartwright@csic.es}
\author{Oreste Piro}
\affiliation{Departament de F\'isica, Universitat de les Illes Balears. E-07122 Palma de Mallorca, Spain.}
\author{Pedro A. S\'anchez}
\affiliation{Institute for Computational Physics, Universit\"at Stuttgart. D-70569 Stuttgart, Germany.}
\email{psanchez@icp.uni-stuttgart.de}
\author{Tom\'as Sintes}
\affiliation{Instituto de F\'{\i}sica Interdisciplinar y Sistemas Complejos, IFISC (CSIC-UIB). Universitat de les Illes Balears. E-07122 Palma de Mallorca, Spain.}

\begin{abstract}
We investigate ice polyamorphism in the context of the two-dimensional Mercedes-Benz model of water. We find a first-order phase transition between a crystalline phase and a high-density amorphous phase. Furthermore we find a reversible transformation between two amorphous structures of high and low density; however we find this to be a continuous and not an abrupt transition, as the low-density amorphous phase does not show structural stability. We discuss the origin of this behavior and its implications with regard to the minimal generic modeling of polyamorphism.
\end{abstract}

\maketitle

\section{Introduction}
Water is one of the most studied substances in all its phases, vapor, liquid and solid --- ice \cite{2012-bartels-rausch} --- due to its ubiquity in nature and great relevance to mankind. Despite the apparent simplicity of this molecule, it shows complex behavior and some of its properties remain poorly understood. Water's hydrogen-bonding and proton-disorder effects lead to a complex phase diagram, which has been progressively extended over many years. An extensive range of crystalline solid phases --- or ice polymorphs --- are known, most of which are stable and/or metastable under extreme conditions. On the other hand, different amorphous solid phases of water, or polyamorphs, have been discovered, including some ices which are identified as the most common water phases in the universe, being those found in interstellar space.

Ice polyamorphs are usually distinguished by their characteristic densities. A low-density amorphous ice (LDA) was first synthesized in the 1930s by physical vapor deposition \cite{1935-burton} and, more recently, by fast cooling of liquid water \cite{1980-bruggeller}. The existence of a second amorphous solid phase of high density (HDA) was established by Mishima and co-workers \cite{1984-mishima} in their experiments on the abrupt pressure-induced amorphization of hexagonal ice (Ih; the crystalline solid phase stable under ambient conditions) at low temperatures. After this discovery, many experimental and theoretical research efforts have been addressed to the characterization of the structural transitions between different crystalline and amorphous solid phases of water \cite{1989-hemley}. Diverse studies suggested the existence of two different mechanisms for the first-order pressure-induced transformation of Ih into HDA: at moderately low temperatures, the amorphization takes place by means of an endothermic melting of the crystalline morphology, whereas at very low temperatures the amorphous phase is the result of an exothermic structural collapse of the crystal \cite{1992-tse, 1996-mishima, 1999-tse}. A reversible pressure-induced transformation of LDA into HDA was also first announced by Mishima \cite{1994-mishima} and subsequently investigated in numerous experimental \cite{2009-winkel, 2010-koza, 2011-winkel} and theoretical \cite{2003-giovambattista, 2003-guillot, 2007-yan, 2008-wang, 2011-hoshino} studies. More recently, the existence of another amorphous solid phase with a very high density (VHDA) has been uncovered \cite{1996-mishima, 2001-loerting, 2005-giovambattista2} and has become the subject of much research \cite{2004-guthrie, 2005-he, 2008-paschek, 2008-koza}.

Studies on amorphous solid water have played a central role in the great interest on the understanding of the phenomenon of polyamorphism --- the existence of different well-defined amorphous phases of the same substance --- that has arisen in recent years \cite{2006-wilding, 2006-loerting, 2008-winkel, 2009-malenkov, 2010-mishima, 2011-loerting}. Solid water polyamorphism has stimulated the search for its liquid counterpart, i.e., the search for two different liquids separated by a first-order phase transition and an associated liquid--liquid critical point \cite{1992-poole, 1997-stanley, 2000-mishima, 2010-mishima}. In this context, a extremely simple physical mechanism has been proposed as a generic explanation for the phenomenon of solid and liquid polyamorphism: the existence of a double well --- or, more generically, of two characteristic length scales --- in the intermolecular potential of a polyamorphic substance \cite{1998-mishima, 2001-franzese, 2005-yan, 2011-vilaseca}.

The extraordinary complexity of water's behavior has favored the development of a myriad of models intended for the study of distinct specific properties. Investigations on water polyamorphism in particular have taken great advantage of computer simulations based on multiple water models with very different levels of detail. For instance, there exist numerous studies on pressure-induced amorphization and other related amorphous transitions performed by molecular-dynamics simulations with atomistic water models \cite{2004-martonak, 2005-martonak, 2005-shun-le, 2011-martonak}. However, an adequate understanding of the essential mechanisms of water polyamorphism --- like the validity of the two-length-scales hypothesis --- may require the exploration of more simple or even minimal modeling approaches. Simple water models have been used for many years to study, for instance, its numerous thermodynamic anomalies and its structural order, phase diagram or solvation properties \cite{1997-nezbeda, 2003-buzano, 2004-pretti, 2005-pretti, 2005-dill, 2005-lishchuk}. Some simple models for water polyamorphism have also been developed \cite{2010-pagnani, 2011-xu}. On the other hand, there exist many simple generic models to test the validity of the two-length-scales hypothesis, mostly based on isotropic central potentials \cite{1998-sadr, 2001-jagla, 2002-wilding, 2002-buldyrev, 2007-franzese, 2011-vilaseca}. Models with anisotropic interactions are far more scarce due to the added complexity imposed by the directional bonds \cite{2010-szortyka, 2011-greschek, 2012-melle}. In the case of water, however, the strong impact of the directionality of the hydrogen bond on its properties makes the use of isotropic potentials particularly challenging, imposing a fine tuning of the model parameters in order to reproduce the desired properties \cite{2008-yan, 2011-abraham}

Perhaps the simplest model of water that incorporates a directional bonding scheme was introduced by Ben-Naim in the early 1970s to obtain a qualitative representation of the open hydrogen-bonded network of molecules that makes up liquid water \cite{1971-ben-naim, 1972-ben-naim}. The model represents water molecules as two-dimensional Lennard-Jones disks with three equivalent hydrogen-bonding arms disposed at 120 degrees, as shown in Figure \ref{fig:model}. The similarity of the shape of these simplified water molecules with a well known brand logotype has led to the adoption of the name ``Mercedes-Benz'' (MB) for the model.

Despite its simplicity, Ben-Naim's MB model and its successive extensions and improvements have been shown to reproduce qualitatively different properties of water, including some of its anomalies and the thermodynamic behavior of the melting transition \cite{1998-silverstein, 1998-silverstein-fpe, 2009-dias}. Mercedes-Benz models have been used to study solvation and hydrophobicity problems \cite{1998-silverstein, 2002-hribar, 2002-urbic, 2011-dias} and the properties of water under confinement \cite{2006-urbic}. Due to its flexibility, this class of models has been the subject of extensive analytical studies \cite{2000-urbic, 2007-urbic, 2009-bizjak, 2010-urbic}. On the other hand and to the best of our knowledge, other more challenging characteristics of water, like the long disputed existence of two liquid phases or the properties of the solid amorphous phases and transitions, have not been explored to date in the minimal MB model. Regarding the main goal of our study, we consider that the MB model may be a useful anisotropic minimal modeling approach to study the essential mechanisms of water polyamorphism. In addition, bond-bending forces play a key role in many low-dimensional systems, such as in the amorphous freezing of soft polymer coils or silica nanoparticles in Langmuir monolayers \cite{2010-maestro}. In water, the interplay between the highly directional hydrogen-bonding network and the geometrical constraints determines the structure of liquid water and ice in two-dimensional layers, either under confinement or at open interfaces \cite{2002-janiak, 2009-lombardo}. Within this context, most computational studies have been devoted so far to the liquid structures \cite{2004-zangi, 2010-fayer}, disregarding the behavior of amorphous solid phases.

In summary, in this work we study for the first time the amorphous solid phases of the two-dimensional MB model of water and their transitions. In particular, we search for the existence of low-density (LDA) and high-density (HDA) solid amorphs and the determination of the nature of the transition between either hexagonal ice (Ih) and HDA as well as between LDA and HDA. Additionally, this approach allow us to study the validity of the two length scales hypothesis when directional bonds and a low dimensionality are introduced in the system. We place our results in the context of the known experimental results about ice.

\section{Simulation Model}
\label{sec:model}

\begin{figure}
 \centering
 \includegraphics*[width=0.85\figwidth]{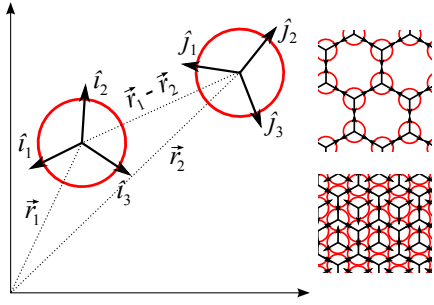}
 \caption{Schematic representation of the two-dimensional Mercedes-Benz model of water (left) and its two crystalline morphologies or polymorphs (right). The water molecules are represented as Lennard-Jones disks with radius $r_{\mathrm{LJ}}$ combined with three hydrogen-bonding arms of length $r_{\mathrm{HB}}$, here depicted as arrows.}
 \label{fig:model}
\end{figure}

The MB pair-interaction potential is expressed as the sum of a radial and a directional term,
\begin{equation}
U_{\mathrm{MB}} \left ( \vec r_i, \vec r_j \right ) = U_{\mathrm{LJ}} \left ( r_{ij} \right ) + U_{\mathrm{HB}} \left ( \vec r_i, \vec r_j \right ).
\end{equation}
The radial term, $U_{\mathrm{LJ}}$, is simply a Lennard-Jones potential,
\begin{equation}
U_{\mathrm{LJ}} ( r ) = 4 \epsilon_{\mathrm{LJ}} \left [ \left ( \frac {\sigma_{\mathrm{LJ}}}{r}  \right )^{12} - \left ( \frac {\sigma_{\mathrm{LJ}}}{r}  \right )^{6} \right ],
\end{equation}
with the distance between the centers of the molecules as argument, $r_{ij} = | \vec r_i - \vec r_j |$. The directional term, $U_{\mathrm{HB}}$, represents the water hydrogen bond and is defined by means of unnormalized gaussian functions, $G\left ( r \right )= \exp \left [ -r^2/ 2 \sigma_{\mathrm{HB}}^2 \right ]$. In his original model, Ben-Naim defined $U_{\mathrm{HB}}$ to be
\begin{equation}
U_{\mathrm{HB}} (\vec r_i, \vec r_j) = \epsilon_{\mathrm{HB}}\ G \left ( r_{ij} - r_{\mathrm{HB}}\right )\ B\left ( \vec r_i, \vec r_j \right ),
\label{eq:Uhb}
\end{equation}
where $\epsilon_{\mathrm{HB}}$ and $r_{\mathrm{HB}}$ are the depth and position of the bonding potential minimum, respectively, and
\begin{equation}
 B\left ( \vec r_i, \vec r_j \right ) = \sum_{k,l=1}^{3} G \left ( \hat \imath_k \cdot \hat u_{ij} - 1 \right ) G \left ( \hat \jmath_l \cdot \hat u_{ij}+ 1 \right ) .
 \label{eq:original-hb}
\end{equation}
Here $\hat \imath_k$ and $\hat \jmath_l$ are unitary vectors in the direction of every hydrogen-bonding arm of molecules $i$ and $j$, respectively, whereas $\hat u_{ij}=\vec r_{ij}/| \vec r_{ij} |$ is the unitary displacement vector between their centers.

More recently, Silverstein and co-workers \cite {1998-silverstein} proposed a computational simplification of the model, by replacing expression (\ref{eq:original-hb}) by:
\begin{equation}
 B\left ( \vec r_i, \vec r_j \right ) = G \left ( v(i, \hat u _{ij}) - 1 \right ) G \left ( w(j, \hat u _{ij}) + 1 \right ),
\end{equation}
where
\begin{eqnarray}
 v(i, \hat u_{ij} ) &=& \max (\hat \imath_1 \cdot \hat u_{ij}, \hat \imath_2 \cdot \hat u_{ij}, \hat \imath_3 \cdot \hat u_{ij}) \\ w(j, \hat u_{ij} ) &=& \min (\hat \jmath_1 \cdot \hat u_{ij}, \hat \jmath_2 \cdot \hat u_{ij}, \hat \jmath_3 \cdot \hat u_{ij})
\end{eqnarray}

According to the previous definitions, the MB model has two different bonding distances given by the minimum of the Lennard-Jones potential, $r_{\mathrm{LJ}}=2^{1/6} \sigma_{\mathrm{LJ}}$, and the hydrogen bond length, $r_{\mathrm{HB}}$, introduced in Eq. (\ref{eq:Uhb}). As a consequence of these two bonding mechanisms, two crystalline solid morphologies can be found in the model: a low-density hydrogen-bonded honeycomb lattice and a high-density triangular lattice of Lennard-Jones disks, as shown in Figure \ref{fig:model}. In particular, it has been shown by means of Monte Carlo NPT simulations that the melting of the MB honeycomb structure reproduces qualitatively the thermodynamic properties of the melting of ice Ih \cite{1998-silverstein}. Therefore, it is reasonable to expect the existence in the model of at least two solid amorphs with low and high characteristic densities, which eventually could be associated with the low- and high-density amorphous ices.

In order to explore the existence and characteristics of solid amorphs in the two-dimensional MB model, we performed extensive equilibrium Monte Carlo NPT simulations with a system composed with up to 1200 MB particles in a rectangular cell with periodic boundary conditions.  At least 50 independent runs of $2\cdot10^7$ steps were performed for every point using the model parameters proposed in previous works \cite{1998-silverstein}: $\epsilon_{\mathrm{LJ}}=0.1$, $\sigma_{\mathrm{LJ}}=0.7$, $\epsilon_{\mathrm{HB}}=1.0$, $\sigma_{\mathrm{HB}}=0.085$, $r_{HB}=1.0$. After equilibration, measures of the internal energy, volume and structure were taken and averaged over all runs. As usual, the system enthalpy and heat capacity were calculated as:
\begin{eqnarray}
H^* &=& U^* + P^*V^*,\\
C_P^* &=& \frac{C_P}{k_B} = \frac {\langle H^{*2} \rangle - \langle H^* \rangle^2}{{T^*}^2}.
\end{eqnarray}
Here $\langle \dots \rangle$ denotes averages over runs and the parameters are expressed in reduced units, relative to the hydrogen bond parameters: $T^* = k_B T / |\epsilon_{HB}|$, $V^* = V / r_{HB}^2$, $U^*=U/|\epsilon_{HB}|$, $H^*=H/|\epsilon_{HB}|$, $P^*=r_{HB}^2 P / | \epsilon_{HB} |$.

In the next section we present the results obtained from our simulations. In many cases, they correspond to simulations performed under very low temperature conditions. 
Under such circumstances one must be aware of the difficulties of obtaining well equilibrated structures at the transition region using simple NPT Monte Carlo simulations; thus extensive computer work is required to avoid the system being trapped into a local minimum. In addition, the simplicity of the MB model --- as with any minimal model --- makes comparison with the experiments relevant only on a qualitative level.

\section{Results and discussion}

In our simulations, we tried to reproduce the experimental amorphization paths established by Mishima and collaborators in their pioneering works on amorphous ices. In particular, we focus on the amorphization of Ih ice --- which we identify with the honeycomb lattice --- into HDA by compression at very low temperature \cite{1984-mishima} and on the reversible transformation between LDA and HDA ices obtained by compression and decompression with annealing \cite{1994-mishima}. Except for the latter case, we worked well below the melting point of the honeycomb crystal, $T_m^* \sim 0.15$ at $P^*=0.1$ \cite{1998-silverstein}, assuming that the resulting sample structures remain in a solid state or, at least, in a very viscous amorphous phase. We shall discuss this assumption on the basis of the rigidity percolation theory applied to amorphous solids.

\subsection{Pressure-induced amorphization of ice Ih}
\begin{figure}[thp]
 \centering
 \subfigure[]{\includegraphics*[width=0.99\figwidth]{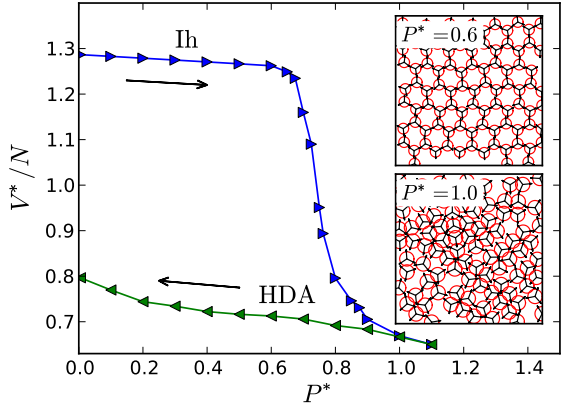}\label{fig:Ih-HDA-volume}}\\
 \subfigure[]{\includegraphics*[width=0.485\figwidth]{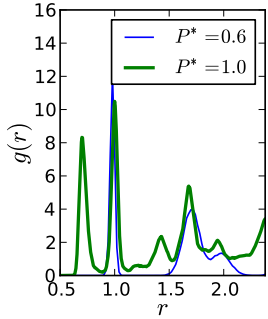}\label{fig:Ih-HDA-rdf}}
 \subfigure[]{\includegraphics*[width=0.5\figwidth]{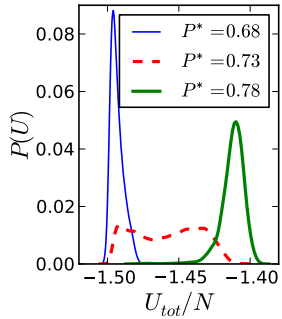}\label{fig:Ih-HDA-histo}}
 \caption{(a) Pressure-induced amorphization of an Ih crystal sample with $N=1200$ at $T^*$=0.05 (upper curve) and decompression of the resulting amorph at the same temperature (lower curve), with insets showing the morphologies found at low and high pressures for the compression curve. (b) From the same sample, radial distribution functions for the stressed Ih crystal ($P^*=0.6$) and for the high-density amorphous phase ($P^*=1.0$). (c) Probability histograms of the configurational energy for three selected pressures of the compression process.}
 \label{fig:Ih-HDA}
\end{figure}

The transformation of ice Ih into HDA is studied by compressing a sample of $N$ molecules, disposed initially in the honeycomb lattice, at a very low temperature, $T^*$=0.05. Figure \ref{fig:Ih-HDA} provides a first insight into the general behavior of the system during this process for $N=1200$ MB particles. As the pressure is increased, the system responds initially with just a slight reduction of the volume and a small displacement of the particles from their equilibrium positions, while keeping the global honeycomb structure. The typical crystalline morphology at $P^*=0.6$ is shown in the inset of Fig. \ref{fig:Ih-HDA-volume}. Consistently, the radial distribution function at $P^*=0.6$ (see Fig. \ref{fig:Ih-HDA-rdf}) identifies two clear maxima corresponding to the first and second nearest neighbor positions in the compressed honeycomb lattice. At around $P^*=0.7$, an abrupt collapse of the stressed honeycomb structure takes place, leading to the arrangement of the particles into a high-density amorphous configuration, which we identify with HDA ice. As revealed by its radial distribution function at $P^*=1.0$ (see Fig. \ref{fig:Ih-HDA-rdf}), this amorphous structure is associated with a significant formation of LJ bonds, corresponding to the peak at around the LJ equilibrium distance, $r_{\mathrm{LJ}}\approx0.78$, which replace a fraction of the original HB bonds of the honeycomb lattice, indicated by the peaks at around $r_{\mathrm{HB}}=1.0$. A snapshot of HDA ice at $P^*=1.0$ is also shown in the inset plot of Fig. \ref{fig:Ih-HDA-volume} to compare with the crystalline structure. The HDA morphology remains with little change when the system is further subjected to an isothermal decompression. Qualitatively, this behavior of the system volume is completely consistent with what can be observed in experiments \cite{1984-mishima} and is a clear indication of a pressure-induced phase transition, probably of a first-order kind, as shown by the abrupt drop in the volume even for such a relatively small system.

We have further investigated the nature of the Ih$\rightarrow$HDA transition by studying different parameters. The analysis of the probability distribution function of the total configurational energy (Fig. \ref{fig:Ih-HDA-histo}) clearly identifies single peaks before ($P^*=0.68$) and after ($P^*=0.78$) the transition, whereas for pressures close to the transition point ($P^*=0.73$) two maxima are found. This result is a strong indication of the first-order nature of the transition \cite{1982-fisher, 1996-wilding}.

\begin{figure}[!t]
 \centering
 \includegraphics*[width=0.99\figwidth]{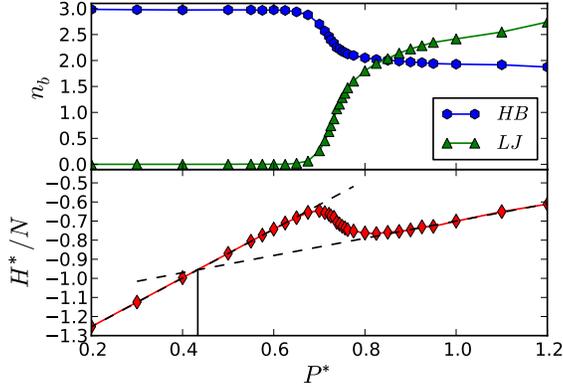}
 \caption{Upper panel: evolution of the mean number of HB and LJ bonds per particle, $n_b$, along the Ih$\rightarrow$HDA transition for $N=1200$; see the text for the bond-counting criterion used. Lower panel: corresponding evolution of the system enthalpy, $H^*/N$, with the projection and intersection of the lines from each side of the transition used to estimate the thermodynamic transition pressure, $P^*_0$.}
 \label{fig:Ih-HDA-enthalpy}
\end{figure}

Another way to characterize this transition is by studying the evolution of the bonds within samples as the pressure is increased. Qualitatively, it is evident that the structure must evolve from a rigid honeycomb lattice, connected by just HB bonds, into a completely different rigid structure, presumably independent from the former in the limit of very high pressures, composed of a triangular lattice of particles in close contact and governed by the soft-core barriers of the LJ potential. In order to obtain some further insight into how this evolution takes place, we computed for every pressure the mean number of bonds of every type and the connectivity of the network defined by all the bonds. The criterion used to take bonds into account has been the following: a bond, either of LJ or HB type, is considered as established between any two given particles when the strength of the interaction is above 0.75 of its maximum possible value. In the case of the LJ potential, since it represents a soft-core barrier for high pressure configurations, we also consider the bond established when the interparticle distance is below its optimum value, $r_{\mathrm{LJ}}$. The upper panel of Figure \ref{fig:Ih-HDA-enthalpy} shows the evolution for the split mean number of HB and LJ bonds obtained for $N=1200$. As expected, there is a sigmoidal-shaped increment of the number of LJ bonds and a reduction of the number of HB bonds in the transition region. The total number of bonds of any kind increases monotonically as one would expect from the different maximum coordination number of both lattices. However, it is remarkable that the number of HB bonds decreases very slowly after the transition region and remains significant at relatively high pressures, indicating that some HB bonds still exist within compact configurations. This behavior has an impact on the thermodynamic properties of the transition. The lower panel of Figure \ref{fig:Ih-HDA-enthalpy} shows the system enthalpy per particle for $N=1200$ as a function of the pressure. The notable step down shown by the enthalpy at the transition region is numerically a consequence of the significant persistence of HB bonds after the collapse of the honeycomb structure, leading to a relatively small increase in the internal energy, $\Delta U^*/N \approx 0.1$, in front of the considerable decrease of the system volume, $\Delta V^* / N \approx 0.5$. Thermodynamically, such drop of enthalpy is a clear indication of the release of a hysteresis heat corresponding to the system relaxation from a metastable state: since the temperature is so low, the system gets kinetically trapped into the crystal phase until the overpressurization is high enough to overcome the energy barriers. Therefore, the thermodynamic transition point can be estimated from the intersection of the projected enthalpy lines from each side of the transition, as shown in Figure \ref{fig:Ih-HDA-enthalpy}. From this calculation we get a value for the transition pressure of $P^*_0=0.43 \pm 0.07$.

Finally, the identification of the bonds allows us to study the clustering of the networks of bonded particles. In all cases we found that the connectivity of the network of bonds is maintained during the transition, so that all the particles remain connected into a single cluster at all pressures. According to the rigidity percolation theory \cite{1998-zallen}, this behavior --- in combination with the monotonic increase in the mean number of bonds --- indicates that the solidity of the sample structure is maintained during the transition. Therefore, this suggests that amorphization occurs via mechanical collapse instead of a melting of the crystal structure.

All these observations are consistent with the known experimental and simulation results on the pressure-induced amorphization of ice Ih at very low temperatures \cite{1992-tse, 1996-mishima, 1999-tse}.

\subsection{Transformations between LDA and HDA ices}

The second process explored in our simulations with the MB model is the reversible transformation between high- and low-density amorphous structures. As in the previous case, we apply a procedure equivalent to Mishima's experiments to simulate a low-density amorphous solid, or LDA ice, and its reversible transformation into HDA \cite{1994-mishima}. Specifically, the transformation HDA$\rightarrow$LDA has been obtained by applying decompression with annealing, i.e., by increasing the temperature of the HDA sample as the pressure is lowered, whereas the reverse transformation HDA$\rightarrow$LDA has been achieved by compressing the LDA ice at high pressure under very low temperature conditions.

\begin{figure}[!ht]
 \centering
 \subfigure[]{\includegraphics*[width=0.98\figwidth]{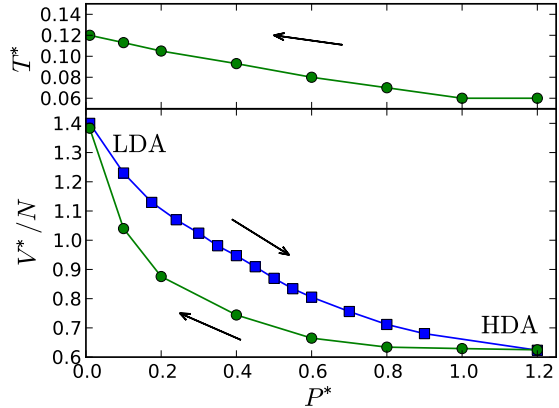}\label{fig:LDA-HDA-volume}}\\
 \subfigure[]{\includegraphics*[width=0.48\figwidth]{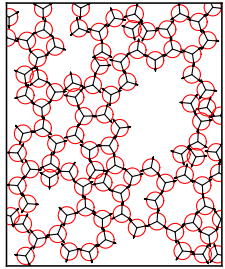}\label{fig:LDA-HDA-config}}
 \subfigure[]{\includegraphics*[width=0.48\figwidth]{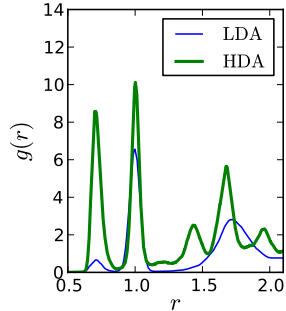}\label{fig:LDA-HDA-RDF}}
 \caption{Results for the reversible transformation between low-density (LDA) and high-density (HDA) amorphous structures by compression and decompression with annealing. (a) Evolution of the system specific volume for the compression at $T^*=0.06$ (upper curve of the lower panel) and decompression with annealing (lower curve of the same panel). The annealing consists of a linear increase of the temperature from $T^*=0.06$ to $T^*=0.13$ (upper panel). (b) Detail of the LDA morphology. (c) Radial distribution functions of the corresponding high-density (HDA) and low-density (LDA) amorphous phases.}
 \label{fig:LDA-HDA}
\end{figure}

Figure \ref{fig:LDA-HDA} shows the main results obtained from the indicated transformation procedures on a sample of size $N=1200$. First, a closed transformation cycle LDA$\leftrightarrows$HDA has been successfully achieved, as shown in the lower panel of Figure \ref{fig:LDA-HDA-volume}. The LDA structure has been produced by a linear increase of the system temperature from 0.05 to 0.13 as the pressure was reduced from 1.2 to 0.01 (upper panel of Figure \ref{fig:LDA-HDA-volume}). The mean number density of the resulting structure, which is mainly controlled by the final temperature, is $\rho^*_{\mathrm{LDA}} \approx 0.71$, a value slightly lower than that corresponding to the honeycomb lattice, $\rho^*_{\mathrm{Ih}} \approx 0.77$. Its radial distribution function, shown in Figure \ref{fig:LDA-HDA-RDF}, confirms that the LDA morphology is the amorphous counterpart of the honeycomb lattice, being mainly composed of HB bonds but with many structural defects. This morphology remains intact when the temperature is set back to $T^*=0.06$. By applying an increasing pressure at such a low temperature, the LDA morphology experiences a gradual compaction to arrive once more at the HDA structure. The continuous, smooth nature of the transformation between these amorphous forms is not what is found in experiments and simulations with other more realistic water models, from which it has been well established that its true nature is that of a first-order phase transition, with associated latent heats \cite{1994-mishima, 2006-loerting-cpc}. We tested also other configurations of potential LDA structures with a somewhat higher density, produced by reducing the maximum temperature of the annealing process. In all cases --- including some with a density even slightly higher than that corresponding to the honeycomb lattice --- the same qualitative results were obtained. For higher maximum annealing temperatures, a complete melting of the structure is soon obtained. Therefore, we were unable to find in the two-dimensional MB model any low-density amorphous structure with enough structural stability to produce a pressure-induced discontinuous phase transition into a high-density amorphous form. Indeed, we want to stress that the transformation cycle shown in Figure \ref{fig:LDA-HDA-volume} corresponds just to the results qualitatively closer to Mishima's experiments that we were able to find in our simulations. In particular, the shape of the decompression HDA$\rightarrow$LDA curve is controlled by the annealing temperatures and therefore can be strongly distorted by using different annealing conditions.

We consider that the origin of the apparent mechanical instability of the low-density amorphous phase in this model is most probably related to the low maximum coordination number imposed by the HB bonds and its interplay with the low dimensionality of the system, which geometrically forbids the existence of defects, inherent to any amorphous structure, without an associated reduction of the mean number of directed bonds. As can be observed in the example of Figure \ref{fig:LDA-HDA-config}, most defects of the LDA structure are associated with misalignments of the directed bonding arms. These misalignments have effects at scales larger than the distance of first-nearest neighbors: as can be observed, the formation of non hexagonal cells --- closed loops of either less or more than six elements --- is very frequent. This effect, combined with the limited possibilities of tessellation of the two-dimensional space, imposes the existence of many unbonded arms. For instance, in the case illustrated by Figure \ref{fig:LDA-HDA}, the total mean coordination number --- calculated by means of the bond-counting criterion introduced in the previous section --- is 2.70, or just 2.43 if only the HB bonds are taken into account. Obviously, any significant decrease of the mean number of HB bonds in this model implies a considerable increase in the total configurational energy of the sample: continuing with the example from Figure \ref{fig:LDA-HDA}, the mean configurational energy per particle of such an LDA structure is approximately -1.34, almost 15\% higher than the energy corresponding to the unstressed honeycomb lattice, -1.57. Such an energy is still significantly higher than that of the stressed honeycomb lattice at the point of collapse, -1.45, showing the overall weakness of the structure. This point represents a significant difference with respect to what is observed in three-dimensional simulations with tetrahedral water models, in which LDA ice keeps the fully coordinated network structure \cite{2011-abraham}.

\section{Concluding remarks}

We have performed extensive NPT Monte Carlo simulations of the two-dimensional MB model in order to study the essential underlying physical mechanisms behind ice polyamorphism. In particular, we have investigated the validity of the two-length-scales hypothesis, previously suggested as the minimal ingredient for the interaction potential of polyamorphic materials, when directional bonds and a low system dimensionality are considered.

To this end we have investigated, in the first place, the pressure-induced transformation of ice Ih into HDA at very low temperatures. Our results suggest the existence of a first-order phase transition in which amorphization occurs via mechanical collapse of the crystal honeycomb lattice from a kinetically trapped metastable state into HDA ice. This amorphous structure is associated with a significant formation of LJ bonds that replace a small fraction of HB bonds in the original crystal. This mechanism ensures the network connectivity during the transition, thus preventing the system from melting. This result is in agreement with the experimental observations of pressure-induced amorphization of ice Ih under very low temperature conditions \cite{1984-mishima, 1996-mishima}.

In the second place we have explored the transformation between high- and low-density amorphous ices by performing an (isothermal compression)--(annealed decompression) cycle. Our results indicate the existence of a continuous transformation between such amorphous structures that is in contradiction with the experimental findings \cite{1994-mishima, 2006-loerting-cpc}. We consider that this discrepancy can be attributed to the low coordination of the low-density amorphous phase, which has no significant structural stability. This low connectivity arises as a consequence of the constraints imposed by the bond directionality and the low dimensionality of the system. Therefore our results provide a clear indication that an effective interaction potential with two characteristic length scales does not guarantee by itself a first-order phase transition between polyamorphs when it is accompanied by strong bonding constraints.

We hope that these results might stimulate new experiments performed in low dimensional systems to study the effect of geometrical constraints and the validity of the predictions of the minimal MB model.

\section*{Acknowledgements}
We thank Itamar Procaccia and Valery Ilyin for introducing us to the MB model and many useful discussions. Simulations were performed at the IFISC's Nuredduna high-throughput computing clusters, supported by the projects GRID-CSIC \cite{GRID-CSIC} and FISICOS (FIS2007-60327, funded by the Spanish MINCNN and the ERDF). JHEC acknowledges MINCINN (Spain) project FIS2010-22322-528C02-02. OP and PAS acknowledge MINCINN (Spain) project FIS2010-22322-528C02-01. TS acknowledges the aforementioned project FISICOS.

\end{document}